# What is Physics?
## The Individual and the Universal, and Seeing Past the Noise

A. R. P. Rau

Department of Physics & Astronomy, Louisiana State University, Baton Rouge, LA 70803, USA

**Abstract**

Along with weaving together observations, experiments, and theoretical constructs into a coherent mesh of understanding of the world around us, physics over its past five centuries has continuously refined the base concepts on which the whole framework is built. This requires looking past obscuring noises present ubiquitously in the real world. The beginnings of physics, classical mechanics with Galileo and Newton, required seeing past friction to get at the basic laws of inertia and motion. Later, in Einstein's Special Theory of Relativity, ideas of location in space and time, and relative velocities in motion, were modified. The recognition that all inertial observers, those moving with constant relative velocity with respect to one another, are on par with respect to a common physics necessitated a re-examination of the concepts of space and time.

Quantum physics, first in non-relativistic mechanics and later in quantum field theories, went further in regarding even familiar concepts of position, momentum, wave or particle, as derived constructs from the classical limit in which we live but not intrinsic to the underlying physics. They are not elements of the underlying reality of the Universe, only the near-universal presence of decoherence making them appear so. It is this noise of decoherence that so shapes our intuition and the very definition of terms with which we experience our world that makes it difficult to understand the quantum nature of that world at its base. Most crucially, the very idea of the individual, whether an object or an event, distinguished only in a mere label of identity from others identical to it in all the physics, exists only as an approximation in what is called the classical limit. It is not an element of underlying quantum reality. Failure to recognize this and seeking alternative explanations in many worlds or multiverses leads only to incoherent logic and incorrect physics.

As an example, in a physical system such as an atom in a particular state, physics deals with the universal system of all such atoms but makes no meaningful prediction of the position of **an** electron or the time of decay of any specific atom. Those are incidental, entirely random among all possible positions and times, even while physics makes very precise predictions for the distribution of the outcomes in measurements on atoms in that state. Physics deals with the universal, not the individual. However, in dealing with large aggregates or degrees of freedom as is most often the case of the classical limit, earlier ideas of position or lifetime are realized as limiting cases.

## Section I.   Introduction

What is physics? To say it is what physicists do, while a truism, does not convey much; it has little content. To say that it is our way of understanding the world we live in, of recognizing underlying principles and laws that connect disparate phenomena of our physical world, has more content and meaning. But it too gives too much weight to **our** understanding of that world rather than to some essence of that world itself. After all, that natural world existed for billions of years before we, or our most ancient hominid ancestors, or life itself, existed on this Earth to question and account for it. Surely, the laws of physics governed the Earth and Heavens throughout that period as well. There existed an underlying physics separate from us just as there is an underlying Universe and reality given to us that we strain to understand.

On the other hand, since our physics is necessarily tied to us, our role as biological beings on Earth that have reached a certain stage of evolution and are guided by concepts and intuition shaped from the cradle on, cannot be completely separated. It is important, therefore, to keep always in mind that our constructs and language and how we represent that underlying reality are a step removed from that reality itself. We are constrained to, and cannot but use, those constructs in our own thinking leave alone in communication with others (and science is 'public knowledge'). But, we have to be able to see past them in grasping the underlying reality. Bohr, a founding father of quantum physics, emphasized this clearly. While we can hope that our models get increasingly more accurate, they always remain models and can only be expected to approach that underlying reality as an asymptote. There is an underlying reality and an underlying physics that are entirely indifferent to our presence on this planet at this particular epoch.

Whatever may be its intrinsic nature, the world around us is seemingly complex. In physics, we often isolate and simplify in the very questions we ask. We then proceed to give sometimes impressively convincing answers, even to great quantitative accuracy. Thus, when a ball is tossed up, among the myriad of questions that may be asked even as regards to its motion, we focus on a few such as how fast was it thrown, how high does it rise, where and when will it fall back onto the ground. The Canadian humorist, Stephen Leacock, describes well the situation that could pertain to our physics questions when he parodies 'word problems' that are taught in middle schools. If Tom digs a hole in one day and Dick digs the same hole in three days while Harry can do so in a half-day, how long would it take to dig the hole when all three work together? Leacock says that even as a school child, he would put up his hand to quiz the teacher on why it should be that they take differing times; in particular, what is wrong with Dick that he is so slow, is he disabled? Or, is he, perhaps, depressed? He says the teacher would ask him to stop acting smart, just sit down, and work out 'the solution.' There is an echo of this in the much later admonition familiar in our times which is attributed to the great quantum physicist Richard Feynman that students of the subject not ask questions of the meaning of quantum physics but just shut up and calculate.

Unfortunately, the opposite has become increasingly fashionable today among some physicists, including some prominent leaders of our field, to go to the other extreme of speculative philosophizing. This ends up being both bad philosophy and bad physics. I list first in Sec. II a few of my criticisms, some as caricatures, to say what physics is not before moving to my perspective in Sec. III on what is the domain of physics, and the nature of a physical system as distinct from its description by an observer, especially one whose intuition has been formed by virtue of being a biological entity in the 'classical limit.' The past five centuries have been a continuous progression of seeing beyond our classical intuition and the very concepts formed by it to the nature of underlying reality and physics.

**Section II. Unfortunate formulations, currently in fashion**

<u>The Many-trajectories Interpretation of a Ball's Motion</u>: Early human experience already recognized that absolute location in space or time of an event has no intrinsic meaning, the origin of coordinates being arbitrary and specific to the observer, typically each regarding his/her own location as the zero. Even the sign, whether left or right, varies with where one sits in a stadium and sees event A of ball being launched and event B of it being caught by a fielder, whether the ball is seen to go from left to right or vice versa. Next, intervals of space and time between A and B seemed more intrinsic and the stuff of physics but, since Einstein, have also been recognized as dependent on the particular inertial frame. Blimps travelling over the stadium with different uniform velocities will ascribe different amounts to the spatial and temporal separation between A and B, only the space-time interval an invariant in common.

In the local 'time-dependent' Newtonian description of the classical mechanics of the ball's motion, it occupies continuously each location in between and is seen to have a trajectory. These are different parabolas as seen by the various stationary and blimp-riding observers. In particular, there is a 'degenerate parabola' of straight up-and-down motion for a blimp travelling with the same horizontal velocity of the ball. The question 'what is the real trajectory of the ball,' this or that parabola or the up-and-down line is, therefore, seen to be 'illegitimate,' not a property of the ball or its motion by itself. The question is incomplete without going on with 'as seen by whom.' It would be an extravagant 'save' of the incomplete question to argue that the ball 'executes' all possible parabolic trajectories, each observer picking one among them. Indeed, in the frame of reference of the ball, it is not the one that moves; rather, it is each observer that performs one of the infinity of parabolas! From this perspective, there is no mystery that there are as many trajectories as there are observers.

Far better, therefore, to spare Occam spinning in his grave, to recognize that the concept of a trajectory is invalid without specifying the observing frame. That is where the infinity lies, in the infinity of possible observers at least in principle. Every observer ascribes a trajectory but it is not an element of the physics of the ball's motion in the sense of relating events A and B. Indeed, in Hamilton's Action Principle formulation of classical mechanics, that physics does not invoke time at all. That is a time-independent description, in which a certain physical quantity called action (wherein time is integrated over) is made stationary. Trajectories are not elements of physics, however real they may seem to our classical intuition, this even in classical physics leave alone for physics in our quantum world. Trajectory as a concept is tied to the observer, and thereby of incidental nature, as are locations specific to each observer, of no general significance. They are not an aspect of the underlying physics.

The above account, entirely within classical physics, mocks The Many Worlds Interpretation of Quantum Physics[1]. The profligate proliferation of worlds and entire universes to say that we, collectively, thread through one set of them reflects again a failure to keep the distinction between an underlying reality and our access to it that necessarily involves us as observers. And, in the case of an underlying quantum reality that appears to be written in **complex** numbers, the very classical concepts of position, time, wave, particle, individual identity, etc., are a mismatch to it. Lacking wave function meters, we bring these concepts and apparatus with which to measure them. And, by definition, outcomes are realized as **real** numbers. We will take this up again in Sec. III.

<u>The Theory of Everything</u>:  Whether or not gravitation can be unified with all the other interactions (a legitimate quest for physics), the 'dream' of an ultimate wrap-up of all laws including also explanations for the observed values of fundamental physical constants in one Theory of Everything[2] seems not only extravagant hubris but to have failed to absorb one of Newton's most fundamental contributions to the nature of our subject. There are laws of physics, such as his of motion and universal gravitation, but he emphasized that there are also 'initial conditions' that are outside of them. A future, more embracing theory may explain these in terms of other entities and laws but there will then be new initial conditions. This may be as much of a profound philosophical contribution of Newton's as are his laws of physics.

Thus, having explained the elliptical nature of any planet's motion, that all in our Solar System lie in a plane was to him an initial condition, an additional feature on which he 'feigned no hypothesis.' Later, Laplace accounted for this in terms of angular momentum conservation in the condensation of an initial large gas cloud, all orbits ending up in the plane of the flattened disk. But, the swirls and initial angular momentum contained in them were new initial conditions. Similarly, it may indeed be that in some future theory, we will derive the value of the fine structure constant that characterizes electromagnetism and thus the structure of our Universe but to expect a closed bootstrap picture with no initial conditions seems untenable. Even in mathematics, such as some specific second order differential equation, solutions are incomplete without specifying initial or boundary conditions that lie outside the domain of the equation itself.

The Anthropic Principle: Closely allied to the above Theory of Everything but more outlandish, and even anti-scientific, are attempts[3] to account for the nature of the world around us and thus of 'physics' from the premise that we are here to ask such questions. Mark Twain, in a 1903 essay[4], already caricatured well this kind of upside-down logic but the attempt fails on its own premises as well. Thus, it is noted, correctly, that from our physics, we can conclude that had the fine-structure constant been different from its observed value, the Universe would have been very different from what we see. Also worthy of note is the sensitivity, that only a few percent change either way makes for a completely different Universe. Granting all this and that with all very young or all very old and dead stars, there would have been no life, biology, and sentient beings to wonder about the nature of the Universe, this can, nevertheless, not 'lead' to a value of 'infinite precision' for the fine structure or any fundamental constant. And we believe that there is such a value, a single precise number, even if increasing experimental accuracy tends towards it only as an asymptote.

The best one can claim from the sensitivity argument is a band of values around the observed one, even if narrow, say of a percent or even lower. Since any narrow band of allowed values still embraces an infinity of irrational numbers, the particular number that is **the** fine structure constant has still not been accounted for. Therefore, anthropic arguments of this kind both fail on their own terms and are fundamentally anti-scientific, closing off prematurely as solved instead of leaving to future quests to gain a more fundamental understanding and accounting. It is falsely declaring victory in explanation instead of accepting, humbly, our inability **at this time** to explain some things, to be left for future science/physics.

<u>Theory, especially when beautiful, suffices</u>: This most recent[5], and perhaps the most problematic, development, ostensibly to defend subjects such as string theory whose experimental confirmation seems to remain forever out of reach, postulates something new to science or physics of accepting theoretical constructs alone and not asking for experimental basis. This compounds the anti-scientific elements of the others above. It forgets that physics is ultimately an experimental subject. No theory, no mathematics, however aesthetically pleasing it may seem to many contemporary particle physicists (they seem to forget that aesthetics, elegance, etc., are subjective), can ask to be excused from being experimentally relevant. The external world is given to us and it is what we are trying to explain. Our imaginations (and our mathematics) are richer, capable of dreaming of and describing multi-dimensional, multi-limbed green-eyed monster inhabited crazy worlds, than the one **actual** world we have been given. That is the one our physics has to account for and it is its existence that acts as a corrective to our mis-steps, keeping our physics honest. That natural world is the final arbiter of truth for our physics. It gives physics a bad name to behave otherwise.

Armchair theorizing even by the best among us cannot be accepted as truth or physics, indeed is a throwback to the pre-Galilean days when it seemed obvious that being at rest was somehow the 'natural' state for physical systems. It was Galileo's careful observation and experimentation that showed the true nature of inertia past the veil of the inescapable 'noise' of friction in the real world around us that brings motions to a halt. The fact that for the underlying physics, it is not zero but uniform velocity that matters, zero as a numerical value (as distinguished from zero or null as a logical or philosophical element) carrying no special status among any other numbers, is what revealed the true nature of inertia, forces, and motion. This marks the beginning of physics.

Again, those who argue for such a drastic re-definition of science or physics, fail on their own terms. An argument is set up that to probe to smaller and smaller distances requires more and more energy and it is simply outside the technological capacity, extrapolated into the future, of our accelerators. So, it is said[5], we should stop requiring such experimental evidence. Several refutations can be given besides the one again of prematurely ruling out breakthroughs in the future simply to declare victory in their own lifetimes (arrogant hubris). Going to smaller distances is set up, parochially, as the most fundamental feature of the subject when one might argue that fundamental laws and principles, regardless of atomic, nuclear or particle scales, are the basis of physics.

Even granting their premises, they seem to forget that complementary variables and approaches may give the same basic physics. Quantum physics in particular has brought to the fore something known since Fourier of the existence of conjugate quantities, the ultra-high in one equivalent to and thus approachable instead through the ultra-low in the other. As an example, parity non-conservation in weak interactions may be observed through high-energy electron-nucleon scattering requiring large accelerators or, alternatively, at low atomic energies by measurements of spectral energy levels to increased resolution. As another example, intermediate states in field theories include all possible particles, even those of 'unreachable mass' for our accelerators, which thus become accessible by going to very highly accurate calculation and experiment at lower energies. In all, one shows only a limitation of one's imagination in blinkered thinking of accelerators of increasing and unaffordable cost and, on that basis, ruling out as unattainable further advances in fundamental physics.

<u>Multiverses</u>: Sharing in common with the above items and the Many Worlds Interpretation, it is fundamentally unscientific, not physics, to invoke other universes[6] with absolutely no link to our own. Everything that has some influence, however weak, on our observed phenomena is, by definition, part of our one Universe and the domain of physics. Other universes as products of our imagination, mathematical or otherwise, which have no connection at all, no possible observation or experiment that would show their presence, are not part of physics. Such invocations of entities are the stuff of fiction, religions, and faith, not of physics or science.

Some even invoke a metaphor of biology to argue for a 'Darwinian selection' among universes for the **right** one with the fundamental constants just so to have led to a world with us in it. This Goldilocks fantasy offends philosophy in multiple ways besides turning upside down, as in Mark Twain's rendition[4], cause and effect. As physics, it fails because of what was noted above that the best that can be claimed is a narrow band of values that may lie around the observed $c$ (speed of light), $h$ (Planck's quantum constant), or $α$ (fine structure constant) but that explains nothing. There is still an infinity of values in that band, leaving that one experimentally observed value unexplained. It also fails as biology, such a teleological argument that sets out to explain a special set of values for a special sentience to ask such questions, being fundamentally antithetical to Darwin's great philosophical contribution. Directed selection, whether by humans or gods, to lead to certain 'desirable' ends is not Darwin's natural selection that admits no goals and would be entirely indifferent to whether a recycling led to a biologically sterile universe, or one with some completely different unimagined biology, or one containing us.

## Section III. The Nature of a Physical System

So, what does physics study? Not an individual particle or event or motion, even if we answer a question of the height reached by a specific ball thrown up just once with a specific speed. No single measurement suffices, as every experimentalist knows while accumulating data from several repetitions. As in science more broadly, multiple balls thrown up multiple times are necessary to build up enough statistics before one confidently predicts what will happen in the single instance. (Much is also abstracted away, the size and shape of balls discarded as irrelevant and, here a point of 'the physics of gravitation,' that even different masses do not matter.) This elementary fact about the experimental method has grown to what has come to be appreciated in quantum physics, that physics does not deal with the individual physical system, individuality being a classical concept valid only in a limit.

Individual identity is a throwback to our classical intuition, not an element of the underlying reality[7]. Everything else being equal, merely slapping on a number as on the backs of runners in a track race, has no place in the quantum description of a physical system where paths and tracks lose meaning. We have to be able to assign values for physical observables so that at least in principle the system is characterized by values of measurable entities. Mere numbering, as if outside of physics and any physical observation or measurement, is indeed just that, outside of and not part of physics. If there is no observable in terms of which one can distinguish between two systems, then no physical meaning attaches to that distinction.

Quantum physics further limits the number of labels because of non-compatibility of physical observables so that there is a maximal set of them (the associated operators mutually commute, for all possible pairings of them). The physical system is characterized by the values for each of that set. Typically, these consist of energy, angular momentum, electric charge, etc., all of which also have absolute conservation laws associated with them. Thus, an elementary particle such as an electron or proton is labeled by its mass (energy), charge, and spin (angular momentum). Mathematically, it is a representation of the Poincare Group with those values for the invariants, experimentally it is registered as such in our detectors as a lump of that much mass $m$, charge $q$, and spin angular momentum $\hbar/2$. One set of these values **is** an electron, another **is** a proton. That is the sense in which we define and use those terms.

An $N$-particle physical system, with numbers from 1 to $N$ labelling them, must always be a superposition of the various permutations among them, symmetric or anti-symmetric under each pairwise interchange of the numbers depending on whether the particles are fermions (half-odd integer spin) or bosons (integer spins). The labels 1 to $N$ themselves are merely for bookkeeping convenience with no intrinsic significance for physics.

Note also nothing is said about the position, in space or time, of an electron so defined. Indeed, physics has progressively shown the limitation of these terms as elements of physics. Thus, as noted, absolute position in space was from the start seen as incidental to the choice of origin. Later, even space intervals, or of time, which were elements of Galilean-Newtonian physics were seen as invalid in Special Relativity as basic quantities. They are specific to a frame of reference, play no role in intrinsic physics. Today, quantum mechanics and quantum field theories relegate them even further. While energy-momentum, as measurable quantities and through their connection to absolute conservation laws are elements of physics, spatial position and time intervals are not. A radioactive nucleus like Radium (Ra) has no position or associated time such as **its** lifetime. Any of the infinity of allowed values, the 'spectrum' of these quantities, is a possible one for a specific Ra nucleus. Any one nucleus singled out may decay into Radon (Rn) plus an alpha particle the next instant or not at all over 'the lifetime of the Universe' (itself also not an intrinsic quantity).

Indeed, the physical system is a Ra nucleus or, even more accurately, the compound object (Rn + alpha particle). A general state is a superposition of the former 'bound' and latter 'continuum' state of that object. (To complicate matters, there may be other possible decay channels as well depending on the isotope involved, which we will ignore for the discussion here.) The same for an atom such as hydrogen, which as a physical system is a (proton + electron) object and, depending on the total energy of the system a superposition of bound and continuum (proton and electron separated to infinite separation) states, extended also to include a photon when coupling to the electromagnetic field is considered (as it always is in the 'real world').

Thus, the ground state of energy (-13.6) eV, assigned the principal quantum number $n$ = 1, is the only true bound state of this system. The state of (p + e + electromagnetic field) of energy (-3.4) eV is the first excited state, bound into that (p + e) configuration in the quantum number $n$ = 2, or the two particles in the ground configuration plus an emitted photon of energy 10.2 eV (Lyman-alpha emission). Being degenerate in energy, the general state is a superposition of the two and it is a matter for the experimental set up on what is observed. If a photon is captured by a detector 'at infinity,' that is, on laboratory scales that are infinitely large relative to atomic size, then it is that we can claim the atom has decayed.

On the other hand, if we surround the system with detectors at infinity and none of them has received a photon, the un-decayed atom is in the $n$ = 2 excited state. In general, with neither of these experimental set-ups and outcomes, what exists is the superposition, one description as a constant flip-flop of Rabi oscillations between the two extremes above. This pertains to a set-up when we surround the system with perfectly reflecting mirrors at some short distance so that no claim can be made of decay having taken place, any incipient photon undergoing reflection by the mirrors to be 'reabsorbed' by the p + e. The same applies to the nuclear radioactivity example although it may be practically more difficult to engineer perfect alpha particle reflectors and re-absorptions.

Thus, again as Bohr clearly recognized and emphasized, the complete apparatus and set-up needs to be specified to define what we are talking about and for us to draw conclusions from physics. To say an unstable entity has decayed has meaning only upon specifying the observation, only when the decay products are observed to have irrevocably separated. That irreversibility is not there intrinsically in quantum physics, arising only upon introducing 'the classical limit,' inevitably brought in by interaction with the classical apparatus (sentience itself has no essential role here). The role of what is called 'decoherence' in that irreversibility will be taken up later but we can already observe that any apparatus involves almost infinitely large number of degrees of freedom and inevitably, therefore, an averaging over many of them.

Therefore, for any radioactive nucleus or excited atom, the very ascribing of a time, a lifetime, is necessarily tied to observation. Any individual excited atom or Ra nucleus can display any number for the value of how long it lived, a lifetime not an intrinsic property of it, but as with the trajectory of a ball in Sec. II, also involving the observation. It is what it is, what is observed, and of no intrinsic significance. For the lifetime of any particular atom or nucleus, all physics provides is the 'spectrum' of allowed values for it, here any number in seconds from zero to infinity.

Only in the aggregate, as with a chunk of Ra containing a large number such as an Avogadro number of nuclei, and referring to the chunk as a single entity, can one regard the lifetime as the time when we will observe half of those nuclei to have become Rn plus alpha particles that have escaped to infinity. This classical limit in which we usually observe, rarely a single Ra nucleus, is what leads to an intuitive but misleading conclusion that lifetime is an intrinsic property of each Ra nucleus. That number characterizes the (Rn + alpha) system, the physics governing that not describing any individual nucleus. The issue here is no different from the naturalness of the concept of the trajectory of a ball, equally misleading, as discussed in Sec. II.

While position and time, as intervals even if not absolute quantities, have no intrinsic significance for a physical system, energy, momentum, and angular momentum do. It is not a coincidence that these are precisely the quantities, along with electric charge, that have the most fundamental laws of physics, the laws of conservation, associated with them. These quantities are conserved in every decay of an atom or nucleus, not just in the aggregate. Indeed, in the early days of quantum physics, even Bohr, Kramers, and Slater[8] toyed with giving up these laws for individual observations while holding them only on average but withdrew against the experimental evidence provided by phenomena such as the Compton Effect and the Franck-Hertz experiment. The laws of conservation are intrinsic elements of physics. And faith in them down to the most microscopic quantum level led to such epochal steps as Pauli's invoking of the neutrino in beta decay to preserve conservation laws. The word faith has been used deliberately and in the sense of science being always provisional knowledge and subject to possible revision with the next experiment. Should one ever show unambiguously a violation of a law of conservation, our physics will need to be modified to accommodate accordingly but, for now, they are among the bedrocks of physics.

As with time or lifetime, a particle's position, location in our three-dimensional space, is also necessarily tied to observation. Absolute space was realized early to depend on the choice of the origin and, therefore, not an element of intrinsic physics. But position in the sense of separation between two points was an item of physics. Indeed, the state of a physical system in Newton's classical mechanics was specified by giving coordinates and velocities of each particle at an instant. Absolute velocities had no meaning but relative velocities were elements of reality. The laws of motion then gave the evolution in time, predicting the state at a later time, once the forces acting were specified. Einstein's Special Theory of Relativity showed the inherent linkage between space and time intervals so that space-time was the element of intrinsic physics of the system, independent of the observer. Each inertial frame's ascribed space and time intervals as separate entities had meaning only to that frame/observer, these numbers being linked by the Lorentz transformation between the frames.

The development of physics has been continuously one of recognizing, and seeing past, the surface masks or veils that hide the underlying reality. Thus, for centuries, it was believed that being at rest (already the implicit caveat, with respect to the Earth!) was the natural state of a physical system. Galileo's experimental observation with inclined planes made him see past the ubiquitous 'noise' of friction to discover the basic nature of inertia, that it is motion with uniform velocity that reveals the intrinsic aspect. Even for the best of his experiments, however, there was always some residual friction that had to be abstracted away to get to the underlying reality only as an asymptote. And with this, zero or any finite velocity was seen as on par, the very basis of Newtonian and later Einsteinian mechanics.

Quantum mechanics took this one step further. Even in its first non-relativistic development, it was clear that both position and velocity could not be simultaneously specified, the two (momentum is used rather than velocity) being incompatible, conjugate objects, with non-zero mutual commutator involving Planck's quantum constant $h$, a fundamental feature of our Universe. Thus, the very definition of the state of a physical system needed drastic modification. One could use either position or momentum, equally, but not both together. Either is a valid 'representation,' capable of giving a complete description of any physical system. Correspondingly, one has a wave function of complex numbers in coordinate or momentum representation.

Along with other representations that we may invoke, these are analogous to the various languages we use in thinking about or communicating with another about our everyday life and experience. The color called 'red,' 'rouge,' 'rot,' etc., has an intrinsic existence beyond these particular references to it. As with the trajectories of a ball in Sec. II, every observer ascribes one of these but they are incidentals, not elements of the underlying reality. So too, are the very concepts of position or linear momentum. They are classical concepts, however intuitive and inevitable they may seem to us. First, just as with language, where dictionaries provide a connection from one language to another, so too unitary transformations (generalization of rotations already familiar in classical physics) provide it in quantum physics, coordinate and momentum being connected by the well-known Fourier transformation. Second, no representation can claim special status or virtue and thus position and momentum are not elements of the underlying reality or physics.

Indeed, had humans encountered quantum physics from the start (which they did not because of the very small value of $h$ on the scale of human experience), coordinate position and momentum may never have been defined as independent concepts. Their Fourier conjugacy would have recognized from the start, that one is a gradient (derivative) in the other space. Once we recognized the universal constant $\hbar$, with precisely the dimensions of the product of momentum $p$ and length $x$, it provided the necessary element to make this connection in full, that $p = -i\hbar(d/dx)$ and, symmetrically, $x = i\hbar(d/dp)$. So also for energy and time, Fourier conjugates with again $\hbar$ the dimensional element of the product of the two dimensions. Time, therefore, is no more than a derivative with respect to energy (as in Wigner's time delay expression in quantum mechanical scattering theory) and is a representation, not an element of intrinsic physics. Whether in classical mechanics in Hamilton's principle of stationary action or in time-independent descriptions of quantum mechanics (even of scattering where the very language that we use seems so tied to a time description), one can do away with the use of space and time to describe physical phenomena. Time-independent scattering theory deals with energy, momentum, phase shifts (which are measurable, indeed related to reflection/transmission coefficients and scattering cross-sections that our experiments measure), all experimentally accessible observables.[9]

Of course, one can equally argue for the primacy of coordinate and time, viewing momentum and energy as their gradients, respectively, all just alternative representations. Because, however, of the very fundamental conservation laws of energy, momentum, and angular momentum, which apply even to individual sub-microscopic phenomena down to the lowest size (correspondingly, highest momentum and energy), one could ascribe more significance to those quantities than to their conjugates that do not have laws of conservation while recognizing that all are representations of an underlying reality independent of them. We will return to this below in a representation-independent abstract description of the state of a physical system.

Seeking a position (or momentum), say of an electron in a hydrogen atom relative to the proton, will of course give a value, just as any observer in the stadium or on a blimp will see a trajectory for the ball's motion (Sec. II) but physics has little to say on this. If a large number of identical copies of the hydrogen atom in exactly the same state are prepared, that measurement of position will give different values on each. Apart from the requirement that these are part of the spectrum of allowed values for position, which are all possible values in three-dimensional space, there is no other prediction from physics, the values realized completely at random. Exactly as with the lifetime of any particular Ra nucleus, there is no significant physics information here. And, as with classical trajectories, the multiple positions seen are because of observations on multiple copies.

With any one copy, were all of space to be covered with electron detectors that fire when a specific lump of $m$, $q$, and spin $s$ that we call an electron is registered, any one of them may fire (some, those at nodes of the wave function, will never fire). It is only for the aggregate that the quantum mechanics of the hydrogen atom in that state becomes relevant, providing the distribution of the detected positions. The same is true for a one- or two- or multiple-slit assembly through which an electron is directed and then received on a screen covered with detectors. Again, in each individual observation, one and only one detector fires but which one is entirely random with no information/physics content. As has been nicely demonstrated experimentally, the corresponding multiple-slit diffraction pattern builds up only in the aggregate. And these patterns are different depending on the number of slits, specific to each set up. On the other side, for any individual measurement of position in a physical system with one electron, the value seen has no information content being the same, a random value, whether the system is a hydrogen atom or a one- or multiple-slit set-up.

Just as it is not that the ball executes an infinity of trajectories (Sec. II), it is not that the electron is simultaneously at more than one point. By definition, an electron is only registered at one detector, all the others then silent, the lump of *m*, *q*, and *s*, not split. Loose statements of the electron being distributed over all space, or depictions of an object, including a narrator (in TV programs aimed at the layman) simultaneously at more than one location, are an aspect of photoshop manipulation, not physics. They mystify and mislead rather than explain the nature of reality. A one-electron system as a matter of definition has only *m* and *q*, different from a system with $2m$ and $2q$ if two electrons were simultaneously detected. That is a completely different system, distinguishable by measurement of charge and mass. There is no mystery, just that position as a very concept inevitably involves an observer and there are as many positions of the electron in that state of the atom as there are observers and copies of that state. Position is not an intrinsic element of underlying reality and, therefore, of physics. The only mystery, or rather discomfort to our intuition, is the acceptance of this new perspective on concepts such as position (or time or momentum). What was more easily accommodated about classical trajectories is a step more difficult for position itself.

The Bohr-Heisenberg philosophy that a physical state be described only by the arrangements of the apparatus that at least in principle describe it, not extraneous (to physics) entities such as mere 'labels of individuality,' was formalized by Dirac in a notation that has come to be universally accepted by physicists. Independent of any representation, the state of a system is indicated by a 'ket' | >, within that bracket symbol standing all the labels measured by those apparatuses. This necessarily restricts to all those physical observables that are mutually compatible, that is, whose operators commute. If this is a maximal set, that is the best that can be used to specify the state. No other labels have physical meaning, in particular mere numbering with integers. Alternative maximal sets are, of course, possible in which case they represent alternative states of the system. Position and linear momentum are often incompatible with energy for many systems such as atoms and slits and thus are not appropriate labels or elements of reality in the sense discussed above.

Thus, the hydrogen atom as a three-dimensional object of an electron and a proton has a maximal set of three labels that may be either the 'spherical' or 'parabolic.' Any state labelled one way is a superposition of many (even infinite) of the other. Among these labels are generally the conserved quantities such as energy and angular momentum and reflect the symmetries that pertain to the set-up. In the absence of external fields, the isotropy of three-dimensional space has besides energy, the total angular momentum and its projection along any axis in space as the three labels constituting the spherical representation. But, if the context is of an external electric field pointing in some direction, only azimuthal symmetry about that axis prevails in place of the full spherical symmetry and another operator replaces the total angular momentum to define the parabolic representation, again with three values. It is, therefore, the experimental context that determines which one we use[10]. Both are equally valid descriptions capable of encompassing all the physics of the hydrogen atom but economy in the number of states used favors one or the other. With an electric field, one or a small handful of parabolic states may suffice whereas an infinity of spherical ones may need to be superposed to describe the same system[11].

More generally as well, such alternative descriptions are ubiquitous in quantum physics. In any scattering experiment, wherein a particle is fired at another (in practice, a beam of particles at an aggregate target), such as an electron at a proton, over most of space when the two have finite separation and a spherically symmetric interaction, angular momentum is conserved. Therefore, the system is viewed in that spherical representation, each value of angular momentum, called a partial wave, independent of and decoupled from others. Only the asymptotic states of preparation and detection, wherein the electron is detected as the lump of $m$, $q$, and $s$ moving in a specific direction and thus with definite linear momentum is more naturally in the parabolic representation. 'Partial wave analysis' involves a passage between the two representations to make contact with the experimental arrangement (this is relevant to footnote 7).

The same applies when we turn from quantities such as position and momentum that have a continuous infinite spectrum to physical observables with a purely discrete and finite spectrum. Most notably, this applies to the intrinsic spin of particles such as an electron or proton, a purely quantum (and relativistic) observable not present in classical physics. Spins do not reside in our three-dimensional space and have no coordinate representation. Unlike other angular momenta associated with rotations in that space, which exist also in the classical limit, these angular momenta admit only representations as matrices or abstract Dirac renderings and have no description in angular coordinates.

A particle with spin $j$ (always an integer or half-odd integer) has ($2j+1$) states and may either be represented by matrices of that dimension or simply as kets $|jm>$, with $m = -j, -j+1, \ldots j-1, j$. In particular, spin $j = 1/2$ has only two states, a 'two-level system,' and experimental meaning attaches only to two possible values, $m = ½, - ½$, that are often called up/down. Electrons and protons have such spin, and being charged particles, can be prepared or detected by arrangements of inhomogeneous magnetic fields called Stern-Gerlach apparatus with two ports at opposite ends. Any orientation of that apparatus will put the particle in one of those ports up/down. That constitutes the entire spectrum, just those two values, and by definition as 'two-level physics,' the only labels and only possible outcomes of measurement for quantum spin.

Again, to avoid loose and confusing argument, it is a matter of being a two-level physical system that any preparation or observation of it, accomplished by a Stern-Gerlach apparatus, has as a matter of definition only two outcomes, up/down. These label the kets of this system. By definition, the Stern-Gerlach apparatus has some orientation in space, only one. That is for a choice we make, of no intrinsic significance just as in the case of assigning a coordinate position or a trajectory for a ball, every observer/observation doing so. There is no question of a single spin $s$ being simultaneously in more than one of them. Two different labels would necessarily mean two spins, taking it outside a one-spin system. A single two-level or spin system does not itself have any orientation, just that it has only two states for any orientation, that orientation in space one imposed by us with our Stern-Gerlach apparatus.

Given the infinity of Stern-Gerlach orientations possible, with unit normal vectors î, these are alternative representations. Since spin projections on different directions do not mutually commute, any one of the two states with respect to some î is a superposition of the two with respect to a different î. And the two states for any orientation are orthogonal, so that being in one excludes being in the other. A famous formulation of Schrödinger dramatized the two-state system such as of a radioactive decay, through a Rube Goldberg amplification, into the states 'dead' and 'alive' of a cat. This has led to confusing mysteries and so-called paradoxes for the layman and even many physicists but that amplification is irrelevant masking for what is involved.

In a two-state system, and if dead and alive are labels for them instead of decayed/not decayed or up/down, every observation is always one or the other with no limbo of something in between dead and alive that the misleading formulations try to steer one to. There are only two states and superpositions of them. There is no third 'limbo,' a superposition not being any such new possibly-observable entity. Notwithstanding all the jazzing up with cats, the physics question is at the simple one degree of freedom with two end points. Until a Stern-Gerlach orientation is picked for preparation or detection, the quantum two-level system is in a superposition for any of the infinity of orientations/representations, again not that it simultaneously possesses all as elements of reality. Indeed, that assumption of simultaneous possession for more than one orientation is false, leading to contradictions ruled out by experiment.

All this is no different than the ball not simultaneously executing an infinity of trajectories in the simple example of Sec. II. Indeed, every Stern-Gerlach observation does yield a result of the system in one port or the other but that is not an element of the underlying reality of the two-state quantum system itself. That resides in some abstract, complex space but it is the Stern-Gerlach apparatus that gives one or the other of two values. Just as every radioactive nucleus is always observed as either intact or decayed into two fragments irrevocably separated, physics making no predictions for it, so too some observations will see the cat dead, others not. That in itself is not mysterious, also true for cats in the real world, that the information becomes available only when the box is opened. If the walls of the cat's cage are perfectly reflecting mirrors, that is all one can say. Only if there is a departure from 100% reflection, the emitted photon or alpha particle escaping to be captured outside the box, can one unambiguously predict a dead cat inside.

Of course, the above is within the premises of the argument as usually posed for the Schrödinger cat, with an assumption that such an immediate mapping from two states of a microscopic atomic or nuclear decay to the states of dead and alive of a cat is possible. That is, of course, a fantastic assumption and, not surprisingly, our usual intuition is shocked. The question of an animal being dead or alive is not one for quantum physics, now or into any distant future. Even within the allied science of biology to which it belongs, we are far from being able to reduce an animal, even the smallest creature, with its many degrees of freedom, to distinguish or define precisely the terms dead and alive. And, surely, the question will never be as simple as being a two-level quantum system. These will be emergent phenomena with meaning only in the classical limit, just as we and many of our apparatus are in that limit.

The very dichotomy between particle and wave, so familiar in everyday experience and throughout classical physics, is not an element of physics or of the underlying reality of our Universe. Often phrased as a 'wave-particle duality,' as if every physical system is simultaneously both, they are again representations, alternative representations through which we access that reality itself which is out of reach. It again requires a precise spelling out of the context including the observing apparatus to see what is manifest. Neither term applies to an electron or any other entity in its quantum reality. It is a state |>, indexed by labels of various physical observables measured on it.

In quantum field theory, an electron may be viewed either in terms of an underlying electron field that exists as a complex continuous function over all space-time, with integrations over all space defining what we measure of its energy, momentum, etc., at any instant. Equally, and completely equivalently, it may be seen as one unit of excitation out of that field, a discrete entity/particle with a certain ($m, q, s$) as captured in our electron detector. In either case, an electron or any other entity is in principle spread over all space-time but in all instances of non-zero $m$, and/or when conserved quantities such as electric charge $q$ or non-integer spin $s$ are involved, we can talk of it as a particle confined to a small position interval in the classical limit. Relativistic quantum physics precludes defining a position to better than the Compton wavelength *h/mc*. Therefore, only entities with *m = 0*, and that too bosons (integer spin) have a wave as a classical limit, all others being particles in that limit. It is no surprise[12] that physicists encountered electrons, protons, and even neutrinos of near negligible $m$ as particles leave alone grains of sand, balls, or ten-ton trucks (all with finite Compton wavelength), and, on the other hand, only electromagnetic radiation as waves.

Where does this classical limit lie and how does it emerge? It is in large part because we ourselves are in that limit and, therefore, likewise are our very concepts, constructs, and apparatus. It is we who impose a mostly-classical description of an underlying quantum world so that it is no surprise that we encounter discomfort in our picture of that world. There is even some irony in that physics is an experimental subject but our observations and experiments involve veils that we must see past to get closer to the essence of underlying reality. We have to see past the obscuring 'noise.' As already noted, relative position in space or time and relative velocities were recognized after The Special Theory of Relativity to be **mere** markers used by any observer to describe motion as a time evolution but not elements of the physics of that motion. All inertial frames observe the same physics even while using different markers. And, some descriptions such as time-independent ones do away altogether with some of them. That all motions in the real world around us are observed to come to a stop, making it seem as if that state of rest is 'natural,' is deceptive and has to be seen through before understanding the nature of inertia, forces, and motion. The very invoking of different inertial frames in relatively constant velocity with respect to each other as on par for physics removes meaning from any fundamental distinction between zero and any other constant velocity.

Next, with the step into quantum physics after recognition of a non-zero ℏ in our Universe, the very concepts of position and its conjugate momentum regarded as independent elements misses the fundamental relationship between them as gradients in each other's conjugate spaces. Even further, non-relativistic quantum mechanics already and today's quantum field theories showed that both position and momentum, previously ascribed as natural elements of particles or waves, were also markers not intrinsic to physics. Three-dimensional space and time provide a background grid against which various quantum fields can be described. Excitations of some energy, momentum, angular momentum, charge, etc., out of vacuum constitute what we observe as particles (if $m$ is non-zero) or waves. Also, from those field operators (Fourier coefficients of expansions of the field functions), one can form combinations that may be seen as quantum-mechanical position and momentum operators. In turn, their spectrums are what we see classically as positions and velocities/momenta.

Just as friction and dissipation obscured a true understanding of inertia before Galileo and Newton, now the very acceptance of position as an intrinsic variable in physics is seen to result from 'decoherence,' that is, the blurring of the many phases that underlie a quantum system[13]. Each degree of freedom is built on complex numbers, any such number realizable as two real quantities of amplitude and phase. The latter is even more delicate to disturbance than are amplitudes subject to decay. Most apparatus that measure physical observables average over most or all these phases. It is in this that the 'classical limit' emerges. Also, amplitude and phase are conjugates, and not simultaneously definable to arbitrary precision except in special circumstances.

All the very concepts of classical physics, whether position of a particle (or its momentum), whether it is a particle or a wave, the use of (mere) labels of identity, are a result of this blurring over or ignoring degrees of freedom. Even in classical physics, all friction and dissipation phenomena have at heart the ignoring of degrees of freedom, energy or flux going into them seemingly lost from those degrees retained. This is compounded in quantum physics. In aggregates of particles, or even with systems of few degrees of freedom when phases get averaged over, one is in the classical limit. (Again, sentience as such has no role; indeed, sentience lies outside the realm of physics.)

Whether for the near universal presence of friction and dissipation distorting the description of motion already in classical physics or for the decoherence of phases in our apparatus defining position, we and our physics need to take heed that there are always a large number of degrees of freedom that are ever present background for even some small sub-system of interest. As with the opening Stephen Leacock item, we isolate for attention only a few but need to be aware of the implicit role played by the rest. However, there is also another crucial aspect throughout all physics of many degrees of freedom. Even when there are many, even infinitely many, such degrees, circumstances may lead to only a few needing consideration. Often associated with the energy scale relevant to the phenomena of interest, degrees of freedom not excited at those energies are ignorable, being frozen out, with the systems in the ground state with respect to those degrees of freedom. Thus, quark degrees of freedom are irrelevant to atomic scale phenomena and to chemistry, and the quantum origin in electron degeneracy pressure of stability of atoms and their aggregates irrelevant in everyday experience and engineering where the hardness of the ground we stand on or of a stone the philosopher kicks can be indexed by macroscopic compressibility indices.

Physics at each level of hierarchy has long been familiar with differing choices of degrees of freedom, even for the very concepts used in our description. Just as even the rocket engineer of today uses non-relativistic Newtonian mechanics, so too through much of physics we continue to use classical-limit terms such as position, momentum, wave, particle, identity, etc. And rightly so, even while holding the recognition that for the underlying reality Special Relativity and quantum fields apply. The very height of a mountain or star is ultimately expressible in terms of fundamental constants such as $\hbar$; indeed, the explanation for their sizes or for that of an atom depends crucially on the value of that and other constants of our relativistic quantum Universe[14]. Nevertheless, it is not necessary or even sensible to go down to the lowest hierarchies when dealing with a higher level. This is not just a practical matter but also conceptual, temperature affording a good example. For a macroscopic fluid, even when recognizing the random thermal kinetic energies of the microscopic atoms and molecules as underlying what our thermometer records when stuck into it, it makes no sense to follow all the Avogadro number of degrees of freedom even were it feasible with computational resources to do so if at the end we seek just one number obtained by averaging out all the rest.

Finally, why is the underlying quantum reality so at odds with the classical limit and our classical concepts and intuition? In large part, it is because of what has been noted about the very delicate nature of the myriad of phases in a quantum system, especially once more than a handful of particles or degrees of freedom are involved. That inevitable blurring at the scale of us and our apparatus accounts in large measure for the way we approach our world through seemingly self-evident concepts such as position, momentum, wave or particle, etc. Future physics may illuminate this further but the mismatch should perhaps also not be so surprising philosophically. It should not shock that the concepts and language used by creatures that have evolved through geology and biology to the current few-hundred year epoch on a small planet in one corner of the Universe may not fit well some language of that underlying reality. Except for hubris, why should we have expected otherwise? Invoking Copernicus, there is nothing special about us for our Universe and its physics.

*Cosmic Landscape: String Theory and the Illusion of Intelligent Design* (Little, Brown, 2005); L. Smolin, Cl. Quant. Grav. **9**, 173 (1992).

[7] Although with a different emphasis and written at a different stage in the development of our subject, Schrödinger's essay on individuality is worth reading in its entirety. Two excerpts relevant to our discussion:
' ... we never experiment with just one electron or atom or (small) molecule. In thought-experiments we sometimes assume that we do; this invariably entails ridiculous consequences as, e.g. that a spherical de Broglie wave, which is supposed to represent 'one' electron, moving in an 'unknown' direction, suddenly collapses into a small wave parcel, when 'that' electron is detected at a definite spot. Nothing of the sort happens if the number 'one' is not ascertained, but may as well be zero or two or three.'
' ... the insight we have now gained into what a particle certainly is *not*; it is *not* a durable little thing with individuality'
E. Schrödinger, British J. Phil. Sc. **3** (11), 233 (1952).

[8] N. Bohr, H. A. Kramers, and J. C. Slater, Phil. Mag. **47**, 785 (1924), Zeit. f. Phys. **24**, 89 (1924).

[9] The conjugacy of time and energy also bears on how the technological capacity of the day makes us approach the quantum world in terms of one or the other of the alternatives. High precision spectroscopy, that is, to better and better energy resolution, of the spectra of atoms in the first decades of the subject are recently giving way to very fast time scale observations that femto- and atto-second laser pulses now provide. Side by side with this development on the experimental side, time-independent theory with expansions in basis states of definite energy are now increasingly replaced by time-dependent wave packet integration that high performance computers make feasible.

[10] An analogous situation pertains to the three flavors of neutrinos and neutrino oscillations. Defining them in terms of their masses ('mass eigenstates') pertains to how they propagate through vacuum or a medium whereas neutrinos are identified experimentally through some interaction in association with an electron, muon, or tau ('flavor eigenstates'), our detectors being sensitive to those or other charged particles coming out of that interaction.

B. Pontecorvo, Sov. Phys. JETP **6**, 429 (1957) and **7**, 172 (1958); Zh. Eksp. Teor. Fiz. **53**, 1717 (1967) [Engl. Transl. Sov. Phys. JETP **26**, 984 (1968)]; Z. Maki, M. Nakagawa, and S. Sakata, Prog. Theor. Phys. **28**, 870 (1962);
https://www.nobelprize.org/nobel_prizes/physics/laureates/2015/advanced-physicsprize2015.pdf

[11] In parallel with footnote 10 about neutrinos, a hydrogen atom, in interacting with the electromagnetic field for emitting or absorbing a quantum of that field is viewed in a spherical basis as appropriate to selection rules (such as electric dipole) but, were a static electric field to pervade all space between such absorptions and emissions, the propagation in between would be described in parabolic terms.

[12] R. E. Peierls, *Surprises in Theoretical Physics* (Princeton U. P., 1979).

[13] M. Schlosshauer, Rev. Mod. Phys. **76**, 1267 (2004).

[14] V. Weisskopf, Science **187**, 605 (1975).